# Starbug fibre positioning robots: performance and reliability enhancements

David M. Brown*, Scott Case, James Gilbert, Michael Goodwin, Daniel Jacobs, Kyler Kuehn, Jon Lawrence, Nuria P. F. Lorente, Vijay Nichani, Will Saunders, Nick Staszak, Julia Tims
Australian Astronomical Observatory, 105 Delhi Rd., North Ryde, NSW 2113, Australia.

## ABSTRACT

Starbugs are miniature piezoelectric 'walking' robots that can be operated in parallel to position many payloads (e.g. optical fibres) across a telescope's focal plane. They consist of two concentric piezo-ceramic tubes that walk with micron step size. In addition to individual optical fibres, Starbugs have moved a payload of 0.75kg at several millimetres per second. The Australian Astronomical Observatory previously developed prototype devices and tested them in the laboratory. Now we are optimising the Starbug design for production and deployment in the TAIPAN instrument, which will be capable of configuring 300 optical fibres over a six degree field-of-view on the UK Schmidt Telescope within a few minutes. The TAIPAN instrument will demonstrate the technology and capability for MANIFEST (Many Instrument Fibre-System) proposed for the Giant Magellan Telescope. Design is addressing: connector density and voltage limitations, mechanical reliability and construction repeatability, field plate residues and scratching, metrology stability, and facilitation of improved motion in all aspects of the design for later evaluation. Here we present the new design features of the AAO TAIPAN Starbug.

## 1. INTRODUCTION

The Starbug prototype [1], which was tested in laboratory conditions in an electrically insulated test frame, consisted of concentric piezo-ceramic tubes with all necessary support wires, vacuum and external metrology fibres temporarily attached (Figure 1). No optical payload (i.e. science fibre) was included. The prototype design has now been reengineered for use in the TAIPAN instrument [2] with a focus on manufacturability, construction repeatability, instrument assembly and maintenance, and ensuring adherence to strict requirements on achievable positioning accuracy.

The new Starbug design contains features including an optical fibre for astronomical observations positioned with high accuracy at the telescope focal plane, and an on-board LED which illuminates three positional metrology fibres. All electrical connections and optical fibres are contained within a vacuum tube, 300mm in length, connecting the Starbug body to a hybrid plug of optical, high and low voltage electrical and vacuum connections (Figure 8). In the telescope, a Starbug adheres by vacuum to a curved glass field plate (GFP) which is located at the telescopes focal plane. The TAIPAN Starbug has been designed with features that ensure the optical clarity of the GFP is minimally compromised.

TAIPAN is a multi-object parallel-positioning fibre-optic spectrograph designed for the UK Schmidt Telescope at Siding Spring Observatory in northern New South Wales, Australia. The instrument will be used to perform galactic and stellar surveys across the whole Southern hemisphere sky, over a 5 year period. In addition to undertaking the TAIPAN surveys, it will serve as a prototype for the MANIFEST fibre positioner system [3] for the future Giant Magellan Telescope. The TAIPAN instrument is designed to use a complement of 300 Starbugs. The initial deployment of the instrument, planned for the end of 2015, will consist of 150 Starbugs. TAIPAN involves significant design in Starbug independent motion control software and electronics. Starbug motion is controlled by 60 processors managing four phases of a 400V, 100Hz waveform that are independently switched to the 2100 Starbug electrodes. A video camera under closed loop control monitors the position of 900 metrology fibres. Software determines the exact location of 300 observing fibres and simultaneously positions them on their respective targets with a tolerance of 5 μm. A Starbug can move on its X and Y axes and can rotate on a point. This allows 300 Starbugs to seek their destination without collision. All aspects of hardware, software and electronic design will incorporate the use of alternate phases of the existing rotation waveforms to increase linear motion speed in future studies.

*david.brown@aao.gov.au Phone +612 9372 4585

## 2. MECHANICAL RELIABILITY AND CONSTRUCTION REPEATABILITY

Several construction challenges with the original test Starbug design have been addressed to improve repeatability of construction and reliability. The piezo tubes (see Figure 1) have electrodes plated both on the inside and outside surfaces. When glue was added to the collar before it was slid into position, the electrodes become contaminated with glue. It was also difficult to achieve repeatable results by applying glue to the tubes and then positioning the collar. It was difficult to achieve a reliable solder joint on the outer tube inner electrode inside a 1mm curved groove that was 1 mm deep. The 2mm length of the joint often failed due to the mechanical lever length of the tubes and the dimensional instability of a piezo active zone.

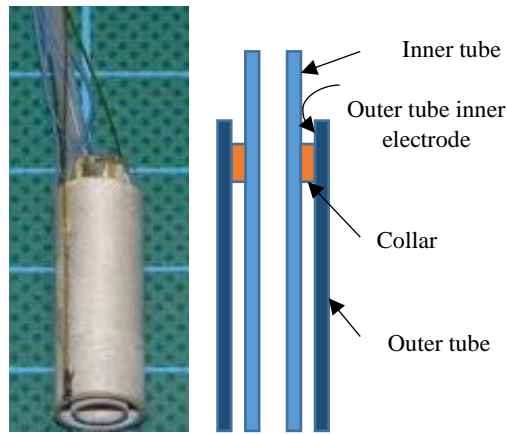

Figure 1; Cross-section of an original test Starbug. All Starbugs mentioned in this paper have an outer piezo outer diameter of 8mm and an inner piezo inner diameter of 4mm.

Design criteria for a production Starbug include: a low centre of gravity (as it requires less vacuum), reduced forces and enhanced protection on the ceramic edges of the piezo tubes, mechanically constrained wires, and increased strength at the interface joints. The piezo tube electrodes have been redesigned without either the inner or outer electrode at the joint to ensure there is no piezo electric effect and no stress on the joint. The joint is increased in length by replacing the collar with a potted joint of low shrinkage and low outgassing epoxy. Potting is achieved by a 3D printed formwork that aligns the tubes and metrology ferrules. It also contains a vacuum throttling hole. Soldering can then be achieved more reliably before assembly. Wires become potted in the joint and 400V arcing is eliminated by its dielectric properties. The potted joint addresses the design criterion to reduce the centre of gravity by allowing the inner tube to be shortened, since soldering tabs beyond the joint are no longer required. A base alignment jig (see Figure 2) also aids alignment and repeatability.

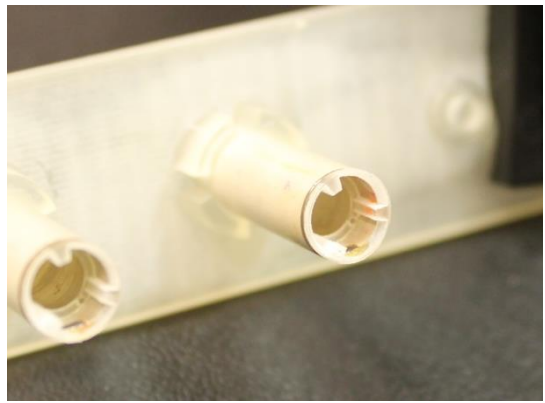

Figure 2; 3D printed formwork on the outer tube before the inner tube is inserted and potting epoxy injected under vacuum. Soldering before assembly on the inner electrode of the outer tube can now be achieved reliably.

# 3. FIELD PLATE CLARITY

## 3.1. Field plate residue and scratching

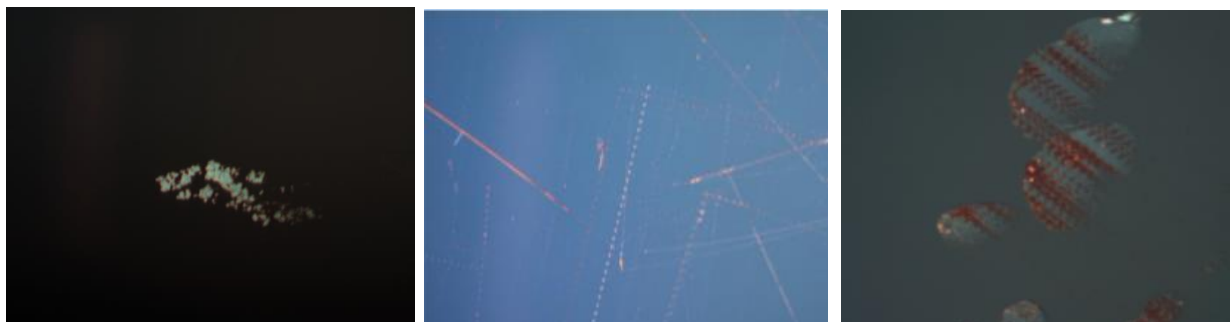

Figure 3;   (a) crumbled ceramic            (b) scratching            (c) trampled finger print oils

A number of residues deposited on the field plate by the prototype Starbug were identified and investigated (Figure 3). The ceramic tubes were found to deteriorate from repetitive use. Deterioration exposes hard materials that are able to scratch the experimental glass field plate. Fingerprint deposits during assembly are effectively ultrasonically cleaned off the ceramic during Starbug motion. Such deposits form globs, fall to the plate and are trampled. New cleanliness procedures have been incorporated in the assembly process including construction in a filtered laminar flow workstation, ultrasonic cleaning of components and tools before and during assembly, wearing gloves and usual clean room handling procedures. Additional procedures have been introduced for storage, protection and ongoing handling of Starbugs to eliminate optical deterioration of the GFP. A waxy residue, erucylamide, was identified by mass spectrometry. It seemed more predominant when the ceramic was freshly polished implying fresh pockets of this substance are exposed to the surface. Frustratingly, the residues persisted after prolonged experiments of activity, vacuum, heat and solvents. Ultrasonic cleaning in solvents caused deterioration of the electrode by flaking and detachment. Outgassing of piezo-ceramic is commonly recommended in vacuum applications. Overnight outgassing did not eliminate erucylamide and prolonged ultrasonic action of the piezo ceramic tubes persisted in shedding residues. These factors lead to experiments of overall tube encapsulation which offered immediate practical results.

## 3.2. Encapsulation

Experiments with nail polishes and urethane coatings were found to eliminate residues, but it is unknown how these coatings would deteriorate over time. Electronic conformal coatings were considered. An electronic conformal coating increases the dielectric strength between electrodes and wires. It also increases electrical safety on the outside. Conformal coatings are designed to ensure continuous coverage and to be durable over time. A conformal coating with UV luminescence has been chosen to provide complete coverage of the entire outer surface area of exposed ceramic or electrode. The effectiveness of the conformal coating for residue containment will be monitored during the life of the TAIPAN instrument.

## 3.3. Slippers – material to separate the ceramic from the glass field plate

A slipper of durable material softer than ceramic or glass is glued to the base of a Starbug (Figure 4) to provide GFP protection, traction, a vacuum seal, mechanical support to the ceramic, and ceramic residue containment on the polished foot of the Starbug. The slipper needs a surface that can be polished to 1 μm or less as a requirement for sufficient vacuum sealing. Materials can be applied that already have a suitable surface finish, but it is difficult to align across both tubes with less than 1 μm flatness, so polishing is necessary.

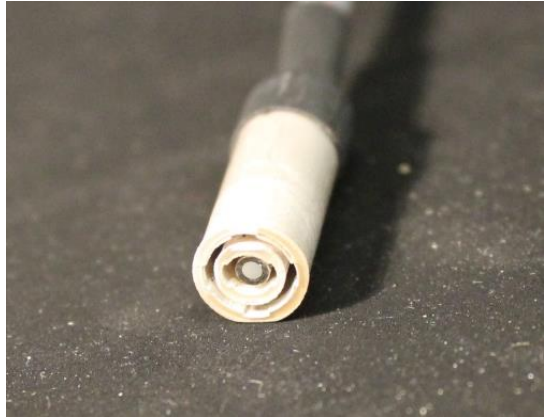

Figure 4; 3D printed slippers are a polished plastic layer glued on the base of the concentric tubes.

3D printing offers features to expedite assembly and introduce added features. Locating tabs make fitting simple and champhers increase predictability of direction. The mechanical properties of coefficients of friction and wearability of 3D printed materials are not available so these properties will be empirically determined.

### 3.4 Vacuum residue

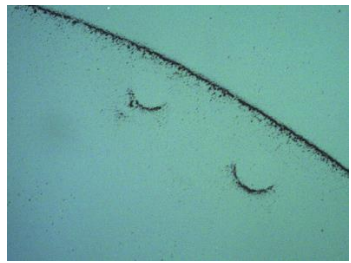

Figure 5; Overnight vacuum residue accumulates on the glass field plate at the 8mm outer diameter edge of a Starbug.

In the TAIPAN scope it was initially envisaged that the Starbugs would remain adhered to the field plate continuously for five years, and a vacuum system would be designed to achieve this. An overnight experiment determined that residues from the air that leak under the Starbug vacuum seal accumulate on the glass field plate as shown in Figure 5. The design intent has been modified so that vacuum is only applied while the instrument is in use (at night). A "bug-catcher" (Figure 6) will be incorporated in the design that positions the Starbugs on the field plate when the vacuum is reapplied.

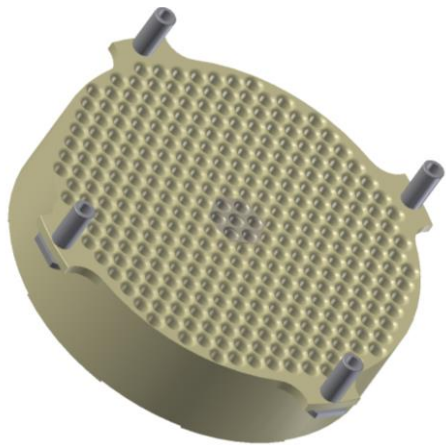

Figure 6; "Bug catcher". The curve of the glass field plate causes the outer Starbugs to tilt off axis, but the bug catcher must travel along the axis, so the diameter of the holes gradually increases from the centre outward. See also Figure 13.

## 4. ADDING PAYLOADS

### 4.1 Astronomy observing fibres and metrology

The TAIPAN requirement for positioning the Starbug has a tolerance of 5 μm. It is reasonable to design in excess of this which may bring future benefits. The prototype Starbug never had a payload or mounted metrology fibres. These are considerably complex issues for stability over five years of use. Design requirements for the observing fibre are: 1) it must be orthogonal to the field plate, 2) the observing fibre face must be within 50 μm of the focal plane, which is 200μm from the field plate, 3) characterization of the positions of the observing and metrology fibres must be consistent for its lifetime.

There is an issue of hysteresis of the piezo ceramic tubes. It can be assumed that when discharged the tubes should be repeatedly mechanically stable, but effectively discharging the tubes in the working environment for final positioning is not practical. There are further issues with attaching metrology fibres to the tube which are the long-term effect of vibration on the fibre and joint, and the stiffening effect of the fibre on the tube. Mounting metrology fibres to piezo tubes is not the most favourable option for maximising reliability and accuracy. The final design mounts everything independently to the joint and nothing impinges on the piezo ceramic tubes, which are then completely free to move.

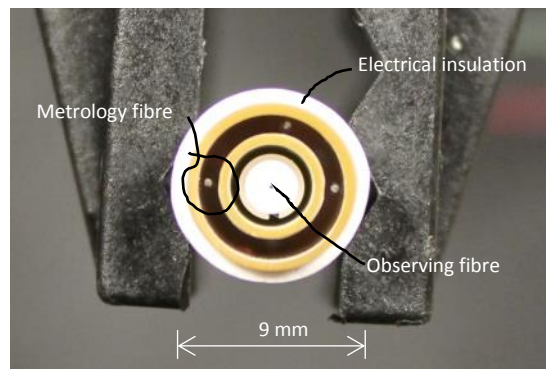

Figure 7; The position of the three metrology fibres is characterised in relation to the central optic fibre for each Starbug.

The astronomical observing fibre is mounted (see Figure 7) in an off-the-shelf ceramic 2.5mm ferrule and associated connection sleeve. This requires 52 N of force to insert so it can be adjusted without gluing. This is useful if the slippers are repolished then the observing fibre can be repositioned at 200μm from the Starbug base. The sleeve is glued to a carbon fibre shaft embedded in the potted piezo tube joint. Due to its dimensions it is absolutely rigid at any orientation so the only positioning error is the 0.1 μm of the metrology fibres, which can be compensated for knowing the telescope angle.

The observing fibre is installed orthogonally by upper and lower alignment jigs. The ceramic tubes are polished in a mechanical fibre polisher with modified holding clamps. When tubes are mounted in the polisher, the alignment of the clamps is made orthogonal by adjustment under an optical measuring device.

Each Starbug has the relative positions of the observing and metrology fibres characterised by back-illumination. This is completed once, but can be verified during its lifespan. The axis of the observing fibre is known in relation to the metrology fibres when they are illuminated during closed loop positioning. Small open loop adjustments will be made during observing integration time to compensate for the rotation of the field.

### 4.2 Guide fibres

Nine guide fibres, each consisting of 1.5mm diameter coherent polymer bundles, are placed evenly across the field plate. They are mounted in Starbugs but go through the connection plate without a connection and are presented at a CCD device behind the connection plate. The guide fibre Starbugs allow object acquisition, telescope guiding, and field plate focus and tip-tilt adjustment. They also provide reference coordinates for the science fibre Starbugs in each observation field. Figure 4 shows a 1.5mm coherent polymer bundle in a Starbug that can be positioned in seconds.

# 5. CONNECTORISATION

## 5.1 Starbug connectors

Prototype Starbugs have been easily damaged in the past, so for this reason it was planned to construct them as a replaceable unit. This required plug connectors for the seven wires, observing fibre plus three metrology fibres and vacuum. This ensures the instrument suffers no downtime due to faulty Starbugs as they can easily be replaced or even upgraded. The plug and socket size is considerably reduced by removing a remotely located metrology LED and placing it on the Starbug body. This replaces three bulky optic fibre connections with two miniature electrical pin connectors. A 15mm diameter hybrid optical fibre / electrical connector as shown in Figure 8 has been sourced that meets the optical focal ratio degradation (FRD) and loss requirements for the TAIPAN instrument. The plug has been modified to accommodate vacuum feed-through and nine wires (Figure 9).

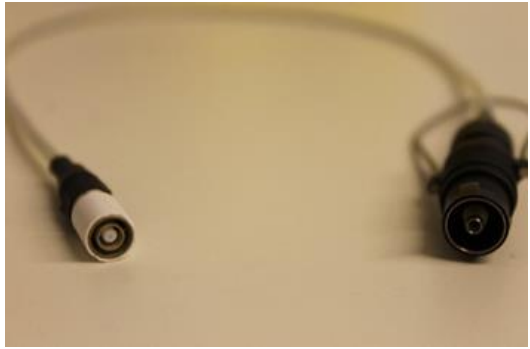

Figure 8; Starbug and connector. Every aspect of design complies with IPC-2221A-Generic Standard on printed circuit board (PCB) design for 400v.

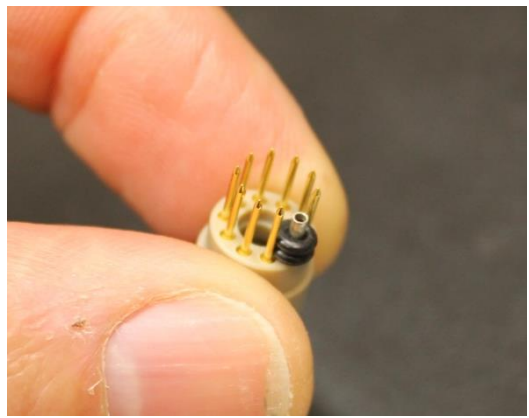

Figure 9; Plug modified for vacuum. Centre hole accommodates an optical fibre connector, not shown.

The resulting density of sockets on the connector plate is hexagonal packing at 17mm centres. This close packing is enabled by eliminating the use of nuts and retaining sockets by a threaded connector plate. The resulting circular 340mm diameter connector plate has 3300 connections consisting of 2700 wire connections, 300 vacuum connections, and 300 observing fibre connections. There is no room for fingers to remove the plugs so the plugs are removed by wire loops that operate a locking mechanism. Behind the connector plate all the cable, vacuum and optic fibres need to be separately connected. Wires are plugged in to PCB's soldered to the Starbug sockets. Connector plug geometric limitations are defined by dielectric requirements of high voltages used, which in turn are defined by trading off heat dissipation with configuration speed. The eventual outcome of reconfiguration trials will determine the consequence of speed on overall reconfiguration time. Figure 10 shows the linear correlation between speed and voltage, indicating voltages above 120v do not significantly increase performance. The highest density plugs available, rated favourably at +/- 177, volts would encircle the connector plate, leaving the centre free for fibres only. An affordable plug is being used which is considerably bulkier (Figure 11).

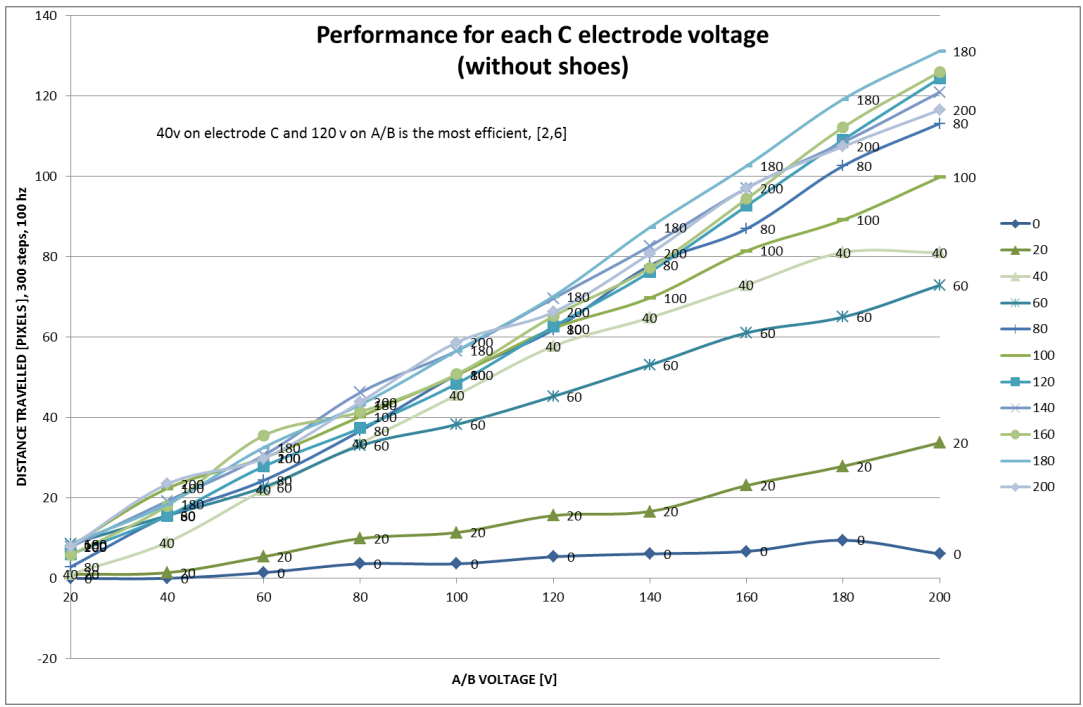

Figure 10; Movement performance due to voltage. Outer electrode voltages (electrode C) above 120V produce consistent performance for all inner electrode voltages (electrodes A & B).

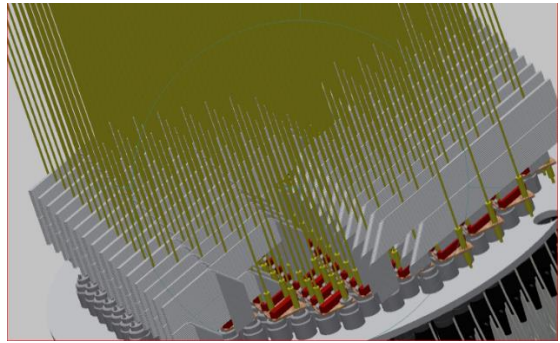

Figure 11; Conventional high density 400V connections (red) connected to grey ribbon cable. Vacuum connections are not shown. Yellow lines are furcation tubes containing astronomical observation fibres.

The mass of the connector plate and Starbugs is 20kg and it moves up and down to place the Starbugs simultaneously on the field plate. The forces on this movement from the flexing of vacuum tubing, 300 science fibres and 2700 wires is going to be measured on a 3D printed model of the connector plate. Science fibre connectors are separately connectable within each plug and socket.

## 5.2 Vacuum manifolds

It would be problematic to have additional vacuum manifolds and connecting tubes amongst the electrical plugs and fibres in Figure 11. One solution being considered is a 3D printed vacuum manifold (Figure 12) that occupies the space between the sockets on the connector plate. The six sections of 50 vacuum outlets could be sequenced with vacuum when the Starbugs are lowered to the glass field plate to enhance adherence. The vacuum system is in the process of being designed. Calculations consider the altitude and temperature range at the UK Schmidt telescope. Each Starbug has a 0.5mm vacuum throttling hole that limits the instrument flow to 250 L/min when all Starbugs are suspended off the glass field plate. One Starbug should adhere with the remaining 299 off the field plate at this flow, followed by an avalanche of adherence that would be expected to complete in milliseconds or seconds. This is yet to be determined empirically, and the time will be used to determine the dimensions of vacuum reservoirs and pump capacities. Either a dummy plug or a working Starbug will be inserted in every socket to control vacuum leakage.

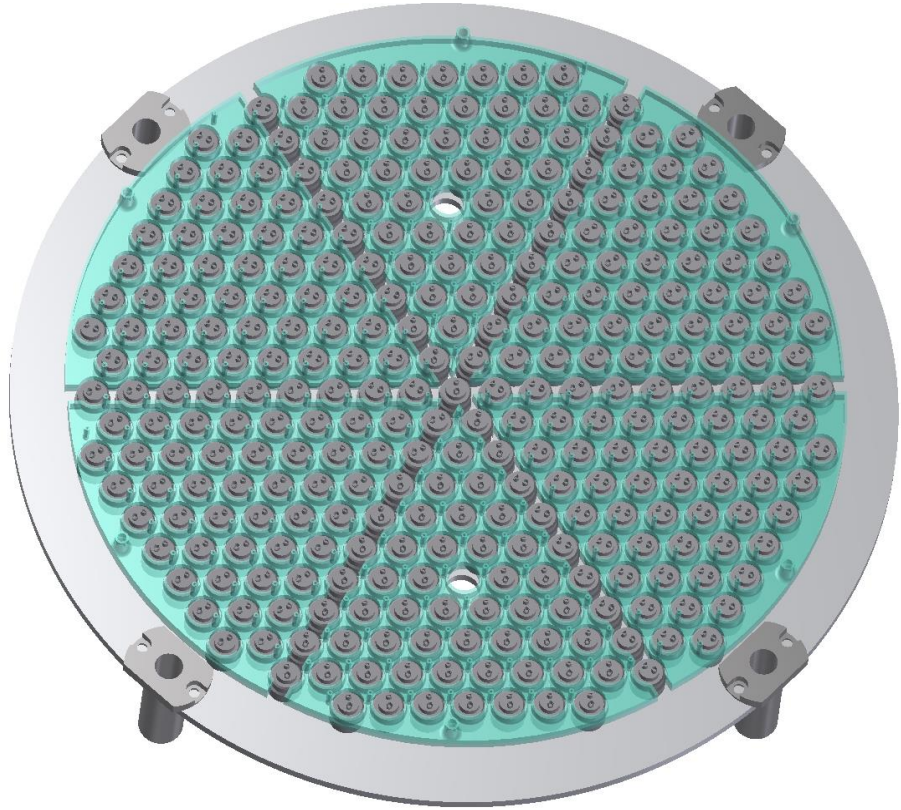

Figure 12: Six wedge-shaped 3-D printed vacuum manifolds (blue), each with 50 vacuum outlets reside under the PCBs (not shown) soldered to the plugs on the connection plate. 300 small silicone tubes (not shown) are connected from the manifold outlets to stainless steel tubes protruding from the sockets.

## 6. STRUCTURE

### 6.1 Design considerations

The instrument structure is currently being designed. The mechanical rigidity of the entire structure must be well within the metrology tolerances. The plane of the glass field plate mount will be aligned to the optical axis of the telescope by three tip tilt motor drives. It can then be moved along the optical axis so the astronomical observing fibres, which are 200 µm within each Starbug, are on the focal plane. The glass field plate can be removed for cleaning and reapplied without realignment. When the telescope is not in use, vacuum is removed and the Starbugs, detached from the glass field plate, sit in the bug catcher. After vacuum is applied, the telescope is positioned at zenith and the connector plate is moved toward the glass field plate so the Starbugs adhere. The bug catcher is then moved against the plugs giving the Starbugs 250mm of working freedom. The connector plate continues to move a further distance so the slack allows the required patrol radius of the Starbugs. The telescope and instrument are now ready for use.

The 3300 connections from the back of the connector plate must be formed so the connector plate can move freely. The bundles must then be formed from the instrument to the telescope wall with a shadow of less than 9mm.

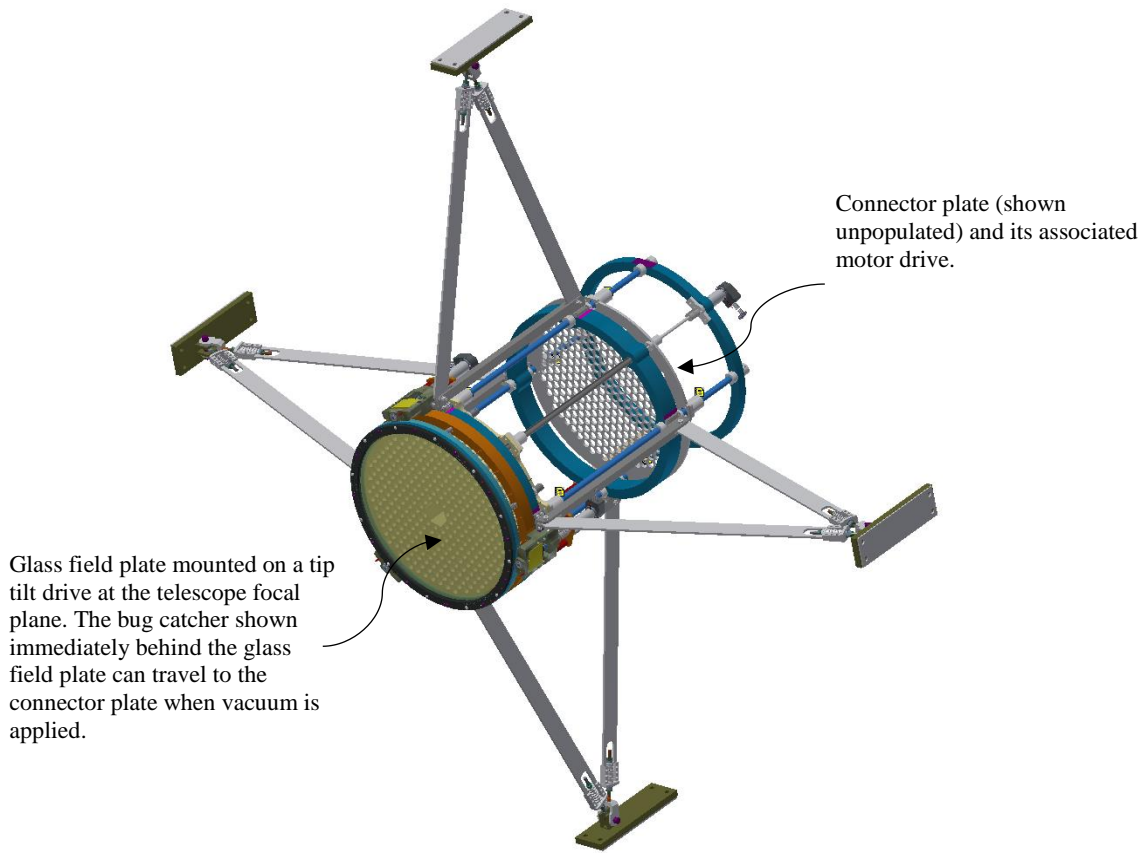

Figure 13; Starbug instrument telescope mount. Not shown is a metrology camera mounted ~3m away in the centre of the main mirror. It faces the base of the Starbugs, and additional fiducial fibres mounted about the glass field plate mount.

### 6.2 Starbug replacement

Access within the telescope is only allowed when the telescope is locked at a near horizontal position. The Starbugs lie in a "bug catcher" which is positioned at the glass field plate. With vacuum applied, there is a working space of 250 mm for inserting or removing plugs from the connector plate. The removal of an individual Starbug requires a special tool that is passed through the 300 cords with a handle protruding out either side. It will have a video display from a fibre bundle looking at the wire loops (see Figure 8) of the plugs. When the Starbug number is identified on the base of the plug, the handles are manoeuvred so a hook engages with the wire loop. At this stage a support structure around the hook will be cradling the plug to accept its weight. The device is pulled so the wire loop releases the locking mechanism in the plug. This removes the vacuum from the Starbug which can then be drawn from the bug catcher. Replacement is the reverse process except the plug is inserted while the Starbug remains outside the instrument. The same tool is then reversed and used to place the Starbug into the bug catcher in a manner likened to endoscopic surgery. A Starbug will be fully testable at the electronics control box outside the telescope.

## 7. CONCLUSION

The design of the AAO Starbug has evolved from an experimental prototype to an optical-payload carrying, durable, reliable and accurate device. Its design encompasses not only its primary purpose, but all events in its lifecycle including construction, storage, connection and maintenance. A complete set of engineering drawings and an assembly procedure is now being finalised for production of 150 instrument-grade Starbugs for the TAIPAN instrument. The five year life of TAIPAN will test the reliability, accuracy, and performance for future projects. In parallel, alternate applications and alternate Starbug payloads are continually being investigated.